# Driving Droplet by Scale Effect on Microstructured Hydrophobic Surfaces

Cunjing Lv[1,2*], Pengfei Hao[2ξ]

[1] Department of Engineering Mechanics, Tsinghua University, Beijing 100084, China

[2] Center for Nano and Micro Mechanics, Tsinghua University, Beijing 100084, China

**Abstract:**

A new type of water droplet transportation on microstructured hydrophobic surface is proposed and investigated experimentally and theoretically ─ water droplet could be driven by scale effect which is different from the traditional methods. Gradient microstructured hydrophobic surface is fabricated in which the area fraction is kept constant, but the scales of the micro-pillars are monotonic changed. When additional water or horizontal vibration is applied, the original water droplet could move unidirectionally to the direction from the small scale to the large scale to decrease its total surface energy. A new mechanism based on line tension model could be used to explain this phenomenon. It is also found that dynamic contact angle decreases with increasing the scale of the micro-pillars along the moving direction. These new findings will deepen our understanding of the relationship between topology and wetting properties, and could be very helpful to design liquid droplet transportation device in microfluidic systems.

## Introduction

The spontaneous motion of liquid on solid surface has attracted great interesting in recently years. For example, micro-fluidic is becoming a hot field and obtaining wide applications, people has developed various methods to realize liquid self-transportatoin.[1-10] So understanding how topology of substrates influence the dynamic behaviors of droplets is essential to clarify the underlying mechanism and practical application.

The main obstacle to droplet motion on a solid surface arises from the hysteresis of contact angles that pin the droplet edge. In order to surmount hysteresis and drive droplet motion, additional energy should be supplied to the droplet in order to produce net force. However, the surface energy of the droplet cannot be converted to mechanical energy spontaneously, so special method should be used, for example, wetting gradient surfaces were prepared by chemical,[3, 4] thermal,[5, 6] electrochemical,[7, 8] photochemical methods[9, 10] and so on.

[*] Electronic address: lvcj05@mails.tsinghua.edu.cn
[ξ] Electronic address: haopf@mail.tsinghua.edu.cn

The possibility of droplet movement due to a surface tension gradient was first predicted by Greenspan[11] and experimentally demonstrated by Chaudhury and Whitesides[12]. Since then people developed various method to realize droplet movement.[13] For example, the direction and velocity of the droplet motion could be manipulated reversibly by varying the direction and steepness of the gradient in light intensity, the reason for the driving force is an imbalance in contact angles generated on both edges of a droplet.[10] When a surface tension gradient is designed into the substrate surface by chemical method, the random movements of droplets were biased toward the more wettable side of the surface.[12] Driven by nonequilibrium noise, periodic motion of droplet was produced along the glass substrates over several tens of seconds.[4] Movements of the droplet could be also controlled by the shape, frequency, and amplitude of the vibration.[14] Moreover, liquid droplet could even move uphill on elaborate designed shape-gradient composite surfaces.[15] Today, the combination of roughness gradient designed substrate and surface tension effect has been applied in micro-pumps for various applications.[1, 2, 16, 17]

The common feature of the above methods is that droplet transportation realized by means of either complex chemical modification or energy import. Here, different from the previous method[1-17], we propose a new way to drive water droplet by scale effect on micropillar-like substrates. We first designed a group of micropillar-like substrates in which the area fractions were kept constant, but the scales of the micro-pillars were decreasing from one side to the other side. When additional water or horizontal vibration was exerted on the original droplet, the droplet could move from the region with small-size micropillars to the rigion with large-size micropillars. To the author's best acknowledge, realizing water droplet transportation depending on scale effect has never been studied. However, this interesting phenomenon could not be explained by the traditional Cassie-Baxter model.[18] Our research demonstrates that scale is very important and should not be ignored especially in small scale.[19-23] The model including line tension not only can be applied to explain these phenomena, but also gives new insights to the wetting theory.

**Driving Droplet by Scale Effect — Experimental Observation**

*Sample Preparation*

The micropillar-like substrates were fabricated by standard photolithography and inductively coupled plasma (ICP) etching techniques, and then a self-assembled monolayer (SAM) of octadecyltrichorosilane (OTS) of formula $C_{18}H_{37}Cl_3Si$ (Acros Organics) was adopted to realize superhydrophobicity by a standard procedure.[24] After the chemical modification, the apparent contact angle on flat surfaces was $105 \pm 1°$.

The apparent static and dynamic contact angles were measured with a commercial contact angle meter (OCAH200, Dataphysics, Germany). The images of the water droplets on the microstructured substrates were observed and recorded by a high-speed CCD, sequential photographs of the wetting behaviors of the water droplets were taken every 10ms. We use a home-made oscillator in which the frequency ranges from 0 to 200Hz, and the amplitudes were measured directly from the experiments.

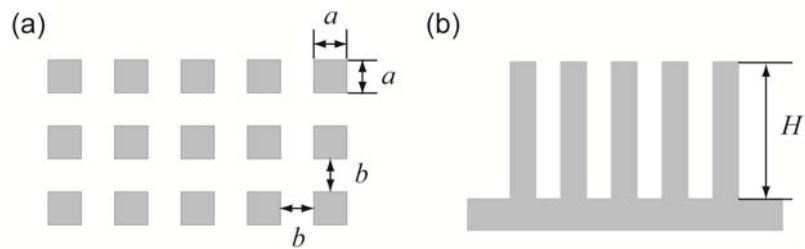

**Figure 1.** Pillar-like micro-structural surfaces: (a) top schematic view of the square pillars; (b) side schematic view of the square pillars

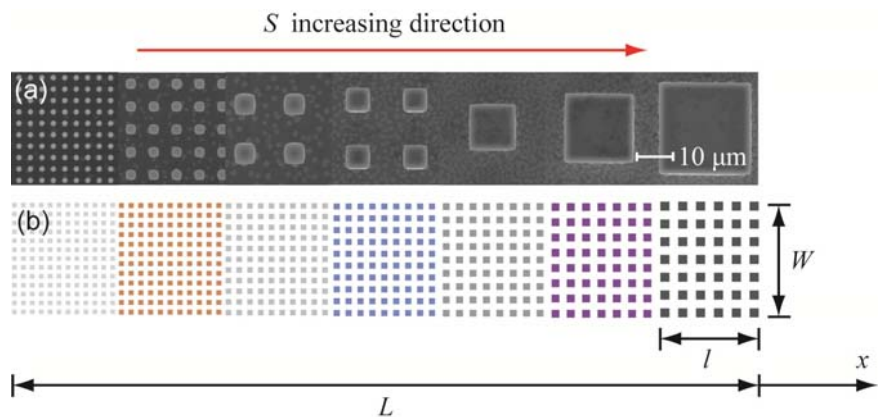

**Figure 2.** Designed parameters of the gradient micropillar-like substrate: (a) SEM images of the gradient substrates, the scale bar is $10\mu m$; (b) Scheme of the gradient substrates with increasing scales in *x* direction, the total width and the length of the substrate is $W=8mm$, $L=14mm$, respectively. The gradient substrate is composed of seven substrates, the width of each single substrate is $l=2mm$.

Different from previous works, in our experiment, a group of micropillar-like substrates were fabricated in which the area fractions were kept constant, but the sizes of the pillar width were decreased from one side to the other side. Figure 2 shows the detailed information of the gradient substrate. Firstly, seven independent small substrates were designed with the width and the length $W = 8\text{mm}$ and $l = 2\text{mm}$, respectively. All of the micropillars had square-shape with height $H \approx 30\mu\text{m}$, and the area frictions were designed at $f = a^2/(a+b)^2 = 0.16$, where $a$ is the side length of the square pillar, $b$ is the spacing between the neighboring pillars. The width $a_i$ $(i = 1,\ldots,7)$ of the micropillars were $2\mu\text{m}$, $4\mu\text{m}$, $6\mu\text{m}$, $10\mu\text{m}$, $20\mu\text{m}$, $30\mu\text{m}$, and $40\mu\text{m}$, respectively. Then the seven substrates were arranged in order of their sizes (Figure 2) and fabricated in a silicon wafer. By this method, we produced a scale-gradient substrate. Here, $S = A/L$ is a shape-dependent roughness scale,[22] given by the boundary length $L$ and area $A$ of the pillar corss-sections. For the above square-shape micropillars, $S_i = a_i/4$.

*Experiment 1: Droplet Motion Realized by Adding Water*

In the first experiment, we will show droplet motion driven by adding water into the original droplet. As shown in Figure 3, in the beginning, a $5\mu\text{L}$ water droplet was first injected from the pinhead, and the substrate was adjusted horizontally in order to ensure the droplet gravity center stand at the middlemost bottom of the pinhead after it was produced. Then, we fixed the substrate, and added $5\mu\text{L}$ additional water into the original droplet one time at the rate of $0.5\mu\text{L/s}$. When the $5\mu\text{L}$ additional water was discharged from the pinhead, there was a short break in order to ensure that the droplet arrive an equilibrium state. Then, $5\mu\text{L}$ additional water was added at the same rate, the circles continued until the droplet became big enough. All the processes were record by a high-speed camera.

Surprisingly, we can see clearly from Figure 3 that the water droplet could always move to the large-scale direction. When the volume of the total droplet is small, this moving behavior seemed not obvious, but when the volume of the droplet was larger than $20\mu\text{L}$, the unidirectional motion of the droplet seemed remarkable.

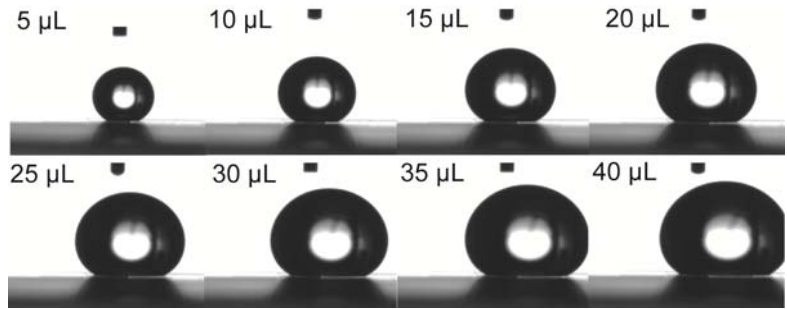

**Figure 3.** Droplet profiles at different positions with increased volume, the position of the pinhead and the substrate were always fixed.

In the above experiment, when the droplet deviated from the middlemost bottom of the pinhead, outflow of additional water might produce hydraulic force and the original droplet might be pushed to move under its horizontal component. In order to make sure that the motion was caused by the scale effect of the micropillar-like substrate, not by the horizontal component of the hydraulic force, we make further experiment below.

When the gravity center of the water droplet moved on the right of the pinhead (Figure 4(d)), we moved the substrate horizontally gently enough to make sure that the gravity center of the water droplet stands on the left of the pinhead (see Figure 4(e)). Then, additional water was continued to add on the droplet. In this case, if the hydraulic force existed, the direction of its horizontal component would be in the opposite direction to the motion of the droplet, and should play a role in resistance. But in the process of dropping, the droplet always moved to the large-scale direction (see Movie 1, Supplementary information). So, we can exclude that the "driven force" pushing the droplet motion was originated from the hydraulic force.

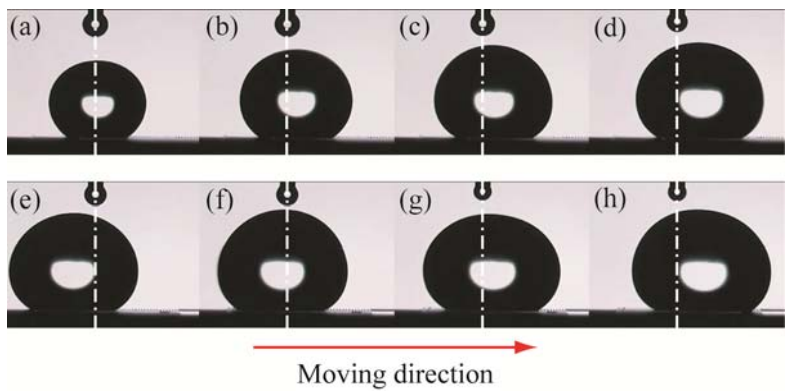

**Figure 4.** Droplet profiles at different positions with sustained increase in volume: (a)(b)(c)(d) the gravity centers of the water droplet were always on the right of the pinhead; (e) the substrate was moved horizontally to let the gravity center of the water droplet stand on the left of the pinhead. The position of the pinhead was always fixed. (see Movie 1, Supplementary Information).

Another interesting phenomenon was that the left and the right apparent contact angles $\theta_L$ and $\theta_R$ on the rear and front of the contact boundary were different from each other not only in the process of the droplet motion, but also in the break of dropping. In order to give detailed information, we plot the relationship between time $t$ and the values of $\theta_L$ and $\theta_R$ (Figure 5). From figure 5 we can see $\theta_L$ and $\theta_R$ exhibiting distinguished features in the process of droplet motion: (i) when the original droplet was disturbed by adding water, fluctuation of $\theta_L$ and $\theta_R$ were around $20°$; (ii) for certain time, $\theta_L$ was always larger than $\theta_R$ which means that the right contact boundary of the droplet (in the forward direction) was always more wettable than that the left contact boundary; (iii) during the 15th second to the 23th second, both of $\theta_L$ and $\theta_R$ increased with increasing of the droplet volume which means gravity could bring influence to the apart contact angle; (iv) during the 23th second to the 43th second, both of $\theta_L$ and $\theta_R$ decreased with time which means the scale of the micropillar was larger, the value of the apparent contact angle is lower.

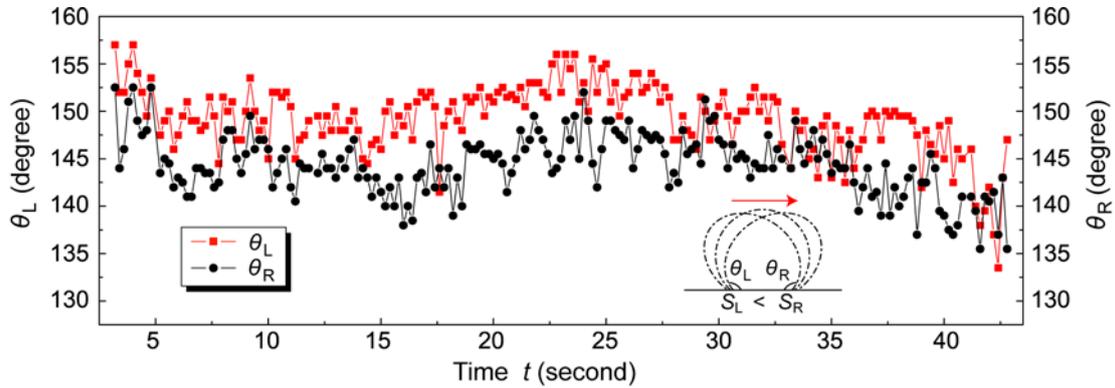

**Figure 5.** Apparent dynamic contact angles $\theta_L$ and $\theta_R$ in the process of droplet motion and in the break of dropping. The red squares represents $\theta_L$, the black circles represents $\theta_R$, respectively.

*Experiment 2: Droplet Motion Realized by Vibration*
In the second experiment, we will show droplet motion driven by mechanical vibration. The substrate was first adhered on a solid platform, and a droplet was released (see Figure 6), we then produced a constant sine horizontal vibration on the platform, and kept the vibration direction parallel to the scale-gradient direction. Interesting, when the volume of the droplet was $20\mu L$, and the frequency and the amplitude of the oscillator were around $80Hz$ and $0.75mm$, the droplet started to move on the same scale-gradient substrate from the small-scale to the large-scale. We tried again for many times, this phenomenon could always happen.

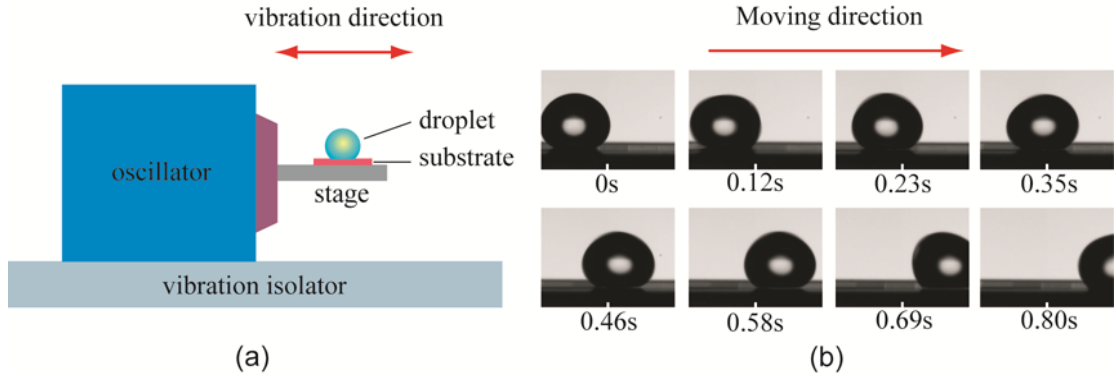

**Figure 6.** Droplet vibration experiment: (a) sketch of the experimental setup; (b) droplet profiles at different positions with constant volume ( $20\mu L$ ) and constant vibration (the frequency and the amplitude of the oscillator were $80Hz$ and $0.75mm$, respectively.), the droplet moved from the small-scale to the large-scale direction. The time bars are shown in the process of the vibration (see Movie 2, Supplementary Information).

In order to give detailed information, we also measured the dynamic contact angles $\theta_L$ and $\theta_R$ varying with time (see Figure 7). Different form Figure 5, $\theta_L$ and $\theta_R$ in Figure 7 are exhibiting distinguished features: (i) when the droplet was disturbed by vibration, the amplitude of the apparent contact angles is larger than Figure 5, and the fluctuation of $\theta_L$ and $\theta_R$ were around $35°$; (ii) the maximum values of $\theta_L$ at different times is as the same or a little larger than the maximum values of $\theta_R$; (iii) and also, the minimum values of $\theta_L$ at different time is as the same or a little larger than the minimum values of $\theta_R$; (vi) in the process of the droplet motion, the trend of the minimum values of $\theta_L$ and $\theta_R$ at different times decreased with the increasing value of the scale of the micro-pillars. The trend of the maximum values of $\theta_L$ and $\theta_R$ at different times also changed, but not so obvious.

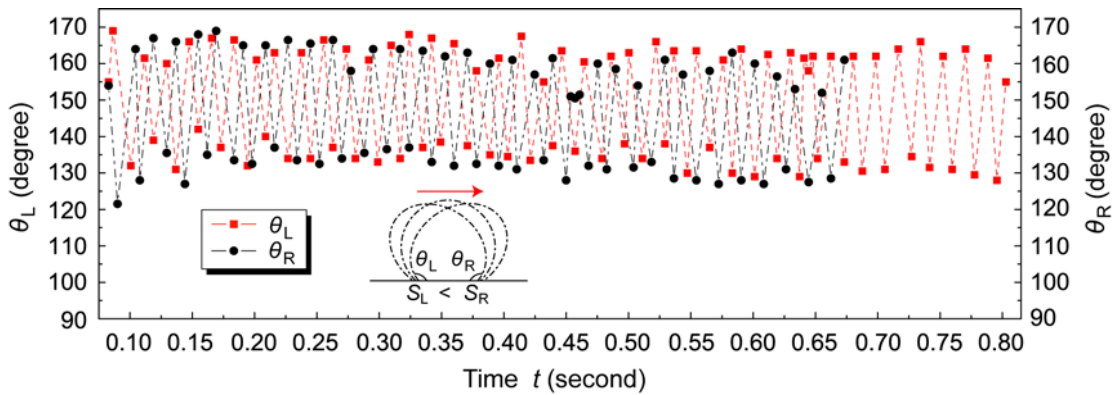

**Figure 7.** Apparent dynamic contact angles $\theta_L$ and $\theta_R$ in the process of the droplet motion. The red squares represent $\theta_L$, the black circles represent $\theta_R$, respectively.

**Theoretical Analysis and Discussion**

As we know, the contact angle $\theta_Y$ reflects wetting characteristic of smooth surface and is given by the well-know Young equation:[25]

$$\cos\theta_Y = \frac{\gamma_{SV} - \gamma_{SL}}{\gamma_{LV}} \tag{1}$$

where $\gamma_{SV}$, $\gamma_{SL}$ and $\gamma_{LV}$ are the surface tension coefficients on solid-vapor, solid-liquid, and liquid-vapor interfaces, respectively.[26] When droplet stands on hydrophobic rough substrate and is in fakir state,[27] the contact areas are composed of water-solid interface and water-vapor interface. The wetting characteristic of such surface was first addressed by Cassie and Baxter,[18] and the apparent contact angle $\theta_C$ was predicted by:

$$\cos\theta_C = -1 + f(\cos\theta_Y + 1) \tag{2}$$

According to Equation (2), we can see that the apparent contact angle $\theta_C$ is only relative with $\theta_Y$ and the area fraction $f$. In other word, once the material systems ($\gamma_{SV}$, $\gamma_{SL}$ and $\gamma_{LV}$) is given, the apparent contact angle $\theta_C$ is only determined by the area fraction $f$. On the contrary, from the above two experiments, the wetting characteristic of the designed gradient substrates was not kept constant even though we fixed the area fraction $f$. So, we conclude that Cassie-Baxter model could not explain wetting properties of water droplet very well on rough surface especially in small scale, because it does not take consideration of the scale effect and the topology of the microstructured substrate.

Actually, Cassie-Baxter equation was deduced from an equivalent energy form on rough surfaces, such as:

$$\cos\theta_C = \frac{\bar{\gamma}_{SV} - \bar{\gamma}_{SL}}{\gamma_{LV}} \tag{3}$$

where $\bar{\gamma}_{SV} = f\gamma_{SV}$ and $\bar{\gamma}_{SL} = (1-f)\gamma_{LV} + f\gamma_{SL}$ are the equivalent solid-vapor surface tension and solid-liquid surface tension, respectively. Put $\bar{\gamma}_{SV}$ and $\bar{\gamma}_{SL}$ into Equation (3), we can easily get Equation (2). According to Cassie-Baxter's idea, for the scale-gradient substrate in our experiment, the equivalent surface tension $\bar{\gamma}_{SV}$ and $\bar{\gamma}_{SL}$ of each small substrate (Figure 2) is the same as each other. While it is not the case, there must be other mechanisms which control the wetting behaviors of water droplet on rough substrate.

Recently, Zheng [22] and Wong [23] recognized the importance of the three-phase contact line tension on the liquid-vapor-solid phase boundary, and they constructed a new model to predict apparent contact angle of water droplet on small-scale rough substrate, independently. Based on their idea, the equivalent solid-liquid surface tension $\tilde{\gamma}_{SL}$ include line tension term can be expressed as:

$$\tilde{\gamma}_{SL} = (1-f)\gamma_{LV} + f\gamma_{SL} + \frac{f\tau}{S} \qquad (4)$$

where $\tau$ is the line tension, and $S=a/4$ for square-shape pillar which reflect the influence of scale effect on wetting properties of rough substrate. Put $\bar{\gamma}_{SV}$ and $\tilde{\gamma}_{SL}$ to Equation (3), we can get Zheng and Wong's model:[22, 23]

$$\cos\theta^* = -1 + f(\cos\theta_Y + 1) - \frac{f\tau}{S\gamma_{LV}} \qquad (5)$$

The value of line tension in Equation (5) was determined to be $\tau = 1.57 \times 10^{-8}$ J/m in Ref. 22. Based on Equation (5), it is not difficult to understand why water droplet could move from the small-scale to the large-scale in the above experiments. If we keep area fraction $f$ constant, $S$ is smaller, the length of the contact line in unit of contact area is longer, so the contribution of the line tension energy is larger, which means the small-scale region is physically more hydrophobic. Even though Equation (5) is applied quantitatively to predict apparent contact angle in Ref. 22, it is still qualitative to compare dynamic apparent contact angles for substrate with different $S$, this conclusion could be validated in Figure 5 and Figure 7. In Figure 5, in the process of droplet motion under disturbance of adding water, $\theta_L$ is almost always larger than $\theta_R$. After the 23th second, both of $\theta_L$ and $\theta_R$ decrease obviously, which means they decrease with increasing the scale of the micropillars. And in this time the contact area is larger enough to occupy substrate with several different $S$. Before the 23th second, this phenomenon is not distinguished, because on one hand the contact area is not large enough to occupy different $S$, on the other hand, the volume of the droplet is too small to withstand the disturbance, so the influence of gravity may larger than the scale effect. In Figure 7, the contact area of the 20μL water droplet is larger enough to occupy two different $S$ or more during vibration. For the minimum value of apparent contact angle at certain time, $\theta_L$ is almost always larger than $\theta_R$. As we know, a lot of previous researches believed that receding contact angle and receding line control the dynamic behaviors of wetting. [24, 28, 29] In our experiment, the minimum value of $\theta_L$ and $\theta_R$ can be treated as receding contact angles of the rear and the front contact lines, so scale-effect is validated further. What's more, the maximum value of $\theta_L$ and $\theta_R$ at a certain time can be treated as advancing contact angles of the rear and the front contact lines. According to some previous researches,[21] advancing contact angle could be very large and have no relationship with the geometrical parameters of the substrate. In Figure 7, most of $\theta_L$ and $\theta_R$ are close to each other, and larger than 165°, and we could not distinguish their difference, our result for the advancing contact angles are consistent with Dorrer's[21] and Zhang's[22] result.

Recently, a lot of researchers recognized the importance of the topology of the microstructure and the three-phase line tension on the wetting properties of the droplet. Chen[19] believed that the topology of the roughness is important for the wetting characteristics. In Öner's experiment[20], the advancing contact angle increased with decreasing of the width of the micro-pillars when the area fractions were kept constant (Table 1 in Ref. 20). In Dorrer's work[21], and Zhang's[30] work the receding contact angle were increased with decreasing of the scale of the micro-pillars when area fractions were kept constant (we can extract this information in Figure 4 and Figure 8 in Ref. 21, and Figure 3 in Ref. 30), but the advancing contact angle seemed have no relationship with the geometrical parameters. Shiu[31] observed experimentally that the contact angle of water droplet increased on various size-reduced polystyrene surfaces. We should emphasize that in Yang's work,[32] the contact angle increased with decreasing of the silica particles sizes on modified silica-coated paper, but what is interesting, the value of contact angle would decrease when the particle size decreased further (Figure 9 in Ref. 31). And later, Zheng's experiment[22] was consistent with Yang's experiment.[33] Recently, Yang[33] studied wetting behaviors of picoliter water droplet on surfaces with grooves of different widths, he not only obtained $-4\times10^{-7}$ J/m and $3\times10^{-5}$ J/m contact line tension on hydrophilic and hydrophobic surfaces, but also observed that that the contact angle increased with increasing of the droplet volume and decreasing of the groove width (see Figure 4 in Ref. 32). Moreover, Raspal's work[34] not only revealed that nanoporous surface may allow the effect on line tension to be visible for micro- to microdroplets, but also emphasized that line-tension model has a physical foundation to solve the contact-angle problem. All the above studies implies the important role of the scale effect for wetting, and this effect is well represented in our experiment.

Based on the above understanding, our experiments and Equation (5) can be used to explain why wetting characteristic could be varied with the topology of the micro-pillar substrates when the area fractions are kept constant.[20, 21, 30, 31, 33] The influence of the line tension is very important for the static and dynamic wetting behaviors of water droplet, and should be took consideration for study the wetting property of rough substrate especially with small scale structures.

Next, we will estimate the value of the driving force on the scale-gradient micropillar structures. Based on Equation (3)–(5), if the scale parameter $S$ is a continuous function of $x$ (see Figure 2), we could give,

$$\frac{\partial}{\partial x}\left(\cos\theta^*\right) = \frac{f\tau}{S^2 \gamma_{LV}} \frac{\partial S}{\partial x} \tag{6}$$

If we ignore the influence of gravity, the liquid-vapor interface of the water droplet will be a spherical surface. The total surface energy could be written as:

$$E = \gamma_{LV}\left(A_{LV} - A_{SL}\cos\theta^*\right) = \frac{3V\gamma_{LV}}{R} \tag{7}$$

where $A_{LV}$, $A_{SL}$, $V$ and $R$ are the area of liquid-vapor interface, solid-liquid interface, volume and radius of the spherical surface, respectively. After some calculations, it not difficult to get the driving force caused by the scale effect along $x$-direction (Figure 2),

$$F_x = -\frac{\partial E}{\partial x} = (3V)^{2/3} \pi^{1/3} \frac{f\tau}{S^2} \frac{(1+\cos\theta^*)}{(1-\cos\theta^*)^{1/3}(2+\cos\theta^*)^{2/3}} \frac{\partial S}{\partial x} \tag{8}$$

Actually, in our experiment, $S$ is not a continuous function of $x$, so we could only estimate the upper and lower limits of $F_x$. Let $S_{12} = (S_2 - S_1)/2$, $\partial S_{12}/\partial x = (S_2 - S_1)/l$, $\theta_{12}^* \approx 160°$, and $S_{67} = (S_7 - S_6)/2$, $\partial S_{67}/\partial x = (S_7 - S_6)/l$, $\theta_{67}^* \approx 150°$ (see Figure 2 and Ref. 22), we can give the upper limit driving force $F_{12} = 1.17 \times 10^{-6}$ N and the lower limit driving force $F_{67} = 9.21 \times 10^{-8}$ N. If let $F_{x\_unit} = F_x/(2R\sin\theta^*)$ (here, $2R\sin\theta^*$ is the width of the solid-liquid contact area), we can get the driving force per unit length: $F_{12\_unit} = 0.001$ N/m and $F_{67\_unit} = 5.45 \times 10^{-5}$ N/m, which is too smaller compared with the liquid-vapor surface tension $\gamma_{LV} = 0.073$ N/m. Because the driving force caused by the scale-effect is small, that is the reason why we should depend on disturbance to help the water droplet to overcome the drag force.

As we know, contact angle hysteresis is the main drag force when droplet movs on solid surface, and the value of contact angle hysteresis is usually estimated as $F_{hys} \approx D\gamma_{LV}\left(\cos\theta_{Rec} - \cos\theta_{Adv}\right)$,[35] here $D$ is the width of the contact area ($D \approx 2R\sin\theta^*$ in our experiments). $\theta_{Rec}$ and $\theta_{Adv}$ are the receding and the advancing contact angles usually defined as the rear and the front contact point on the droplet motion direction. But in Figure 5, $\theta_{Rec}$ are larger than $\theta_{Adv}$, so the water droplet could obtain driving force from the contact angle difference. In Figure 7, severe disturbance happened, anyway, the water droplet could also obtain driving force from the contact angle difference. So, it's interesting in our experiment that the resistance to motion was not caused by the above defined contact angle hysteresis, but might come from friction, adhesion or pinning of the contact line which should be further studied.

**Concluding Remarks**

In this paper, we revealed that the "scale effect" could be considered as a "driving force" to realize water droplet transportation on hydrophobic substrate with scale-gradient microstructures. When vibration or additional water was applied on the original droplet as disturbance, the droplet will always move to the region from the small-scale to the large-scale in order to decrease its total surface energy. During the process of droplet motion, the apparent dynamic contact angle on the small-scale side would be always larger than the large-scale side even though the area fractions were kept constant. However, the traditional Cassie-Baxter model does not take consideration of the influences of topology and scale effect on the wetting properties of rough surface, so it could not be applied to predict apparent contact angle and droplet transportation behaviors in our experiments. Different from the previous understanding, we revealed that the line tension could be considered as the mechanism and should not be ignored especially in small scale.

What's more, our discovery may give new explanations qualitatively why there are so many plants and insects which own excellent hydrophobic/superhydrophobic properties have small-scale micro-nano structures,[36, 37] small-scale could help them to keep away from wetting when external perturbation happens. Our work could guide to design optimal microstructures to realize excellent superhydrophobic properties and special functions in micro-fluidic field for practical applications.

**Acknowledge**

Supports by the Chinese NSFC under Grant No. 10672089 and 10872106 are gratefully acknowledged.

**References**

(1) Luo, J. K.; Fu, Y. Q.; Li, Y.; Du, X, Y.; Walton, A. J.; Milne, W. I. *J. Micromech. Microeng.* **2009**, *19*, 054001,
(2) Nguyen, N. T.; Huang, X.; Chuan, T. K. *J. Fluid Eng.* **2002**, *124*, 384-392.
(3) Bain, C. D.; Burnett-Hall. G. D.; Montgomerie, R. R. *Nature* **1994**, *372*, 414-415.
(4) Sumino, Y.; Magome, N.; Hamada, T.; Yoshikawa, K. *Phys. Rev. Lett.* **2005**, *94*, 068301.
(5) Linke, H.; Aleman, B. J.; Meling, L. D.; Taormina, M. J.; Francis, M. J.; Dow-Hygelund, C. C.;

Narayanan, V.; Taylor, R. P.; Stout, A. *Phys. Rev. Lett.* **2006**, *96*, 154502.

(6) Lagubeau, G.; Merrer, M. L.; Clanet, C.; Quere, D. *Nat. Phys.* **2011**, *7*, 395-398.
(7) Gallardo, B. S.; Gupta, V. K.; Eagerton, F. D.; Jong, L. I.; Craig, V. S.; Shah, R. R.; Abbort, N. L. *Science* **1999**, *283*, 57-60.
(8) Yamada, R.; Tada, H. *Langmuir* **2005**, *31*, 4254-4256.
(9) Abbort, S.; Ralston, J.; Reynolds, G.; Hayes, R. *Langmuir* **1999**, *15*, 8923-8928.
(10) Ichimura, K.; Oh, S.-K.; Nakagawa, M. *Science* **2000**, *288*, 1624-1626.
(11) Greenspan, H. P. *J. Fluid Mech.* **1978**, *84*, 125-143.
(12) Chaudhury, M. K.; Whiteside, G. M. *Science* **1992**, *256*, 1539-1541.
(13) Darhuber, A. A.; Troian, S. M. *Annu. Rev. Fluid Mech.* **2005**, *37*, 425-455.
(14) Genzer, J.; Bhat, R. R. *Langmuir* **2008**, *24*, 2294-2317.
(15) Zhang, J. L.; Han, Y. C. *Langmuir* **2007**, *23*, 6136-6141.
(16) Cheng, C.-M.; Liu, C.-H. *J. Microelectromech. Syst.* **2007**, *16*, 1095-1105.
(17) Yang, J.-T.; Chen, J. C.; Huang, K.-J.; Yeh, J. A. *J. Microelectromech. Syst.* **2006**, *15*, 697-707.
(18) Cassie, A. B. D.; Baxter, S. *Trans. Faraday Soc.* **1944**, *40*, 546-551.
(19) Chen, W.; Fadeev A. Y.; Hsieh M. C.; Oner, D.; Youngblood, J.; McCarthy T. J. *Langmuir* **1999**, *15*, 3395-3399.
(20) Öner, D.; McCarthy, T. J. *Langmuir* **2000**, *16*, 7777-7782.
(21) Dorrer, C.; Rühe, J. *Langmuir* **2006**, *22*, 7652-7657.
(22) Zheng, Q.-S.; Lv C. J.; Hao, P. F.; Sheridan, J. *Science China, Physics, Mechanics & Astronomy* **2010**, *53*, 2245-2259.
(23) Wong, T.-S.; Ho, C.-M. *Langmuir* **2009**, *25*, 12851-12854.
(24) Lv, C. J.; Yang, C. W.; Hao, P. F.; He, F.; Zheng, Q.-S. *Langmuir* **2010**, *26*, 8704-8708.
(25) Young, T. *Phil. Trans. R. Soc. Lond.* **1805**, *95*, 65-87.
(26) de Gennes, P. G.; Brochard-Wyart, F.; Quéré, D. *Capillarity and Wetting Phenomena*, Spring; Berlin, **2003**.
(27) Quéré, D. Fakir droplets, *Nat Mater.* **2002**, *1*, 14-15.
(28) Gao, L.; McCarthy, T. J. *Langmuir* **2006**, *22*, 6234-6237.
(29) Gao, L.; McCarthy, T. J. *Langmuir* **2006**, *22*, 2966-2967.
(30) Zhang, X.; Mi Y. *Langmuir* **2009**, *25*, 3212-3218.
(31) Shiu, J.-Y.; Kuo, C.-W.; Chen, P.; Mou, C.-Y. *Chem. Mater.* **2004**, *16*, 561-564.
(32) Yang, H.; Deng, Y. Journal of Colloid and Interface *Science* **2008**, *325*, 588-593.
(33) Yang, J.; Rose, F.; Gadegaard, N.; Alexander, M. R. *Langmuir* **2009**, *25*, 2567-2571.
(34) Raspal, V.; Awitor K. O.; Massard, C.; Feschet-Chassot, E.; Bokalawela, R. S. P.; Johnson, M. B. *Langmuir* **2012**, *28*, 11064-11071.
(35) Frenkel, Y. I. J. *Exptl. Theoret Phys. (USSR)* **1948**, *18*, 695.
(36) Bush, J. W. M.; Hu, D. L.; Prakash, M. *Advances in Insect Physiology* **2008**, *34*, 117-192.
(37) Guo, Z.; Liu W. *Plant Science*, **2007**, *172*, 1103-1112.

**Supplementary Information**

**Movie. 1** When additional water is adding into the original droplet, it always moves from the small scale to the large scale on the substrate no matter the relative position between the pinhead and the gravity center of the water droplet. The area fraction of this scale-gradient substrate is kept constant at $f = 0.16$.

**Movie. 2** Droplet moves on the scale-gradient substrate in which the area fraction is kept constant at $f = 0.16$, but the scales of the micro-pillars is increasing from the left side to the right side. When steady vibration is produced, the droplet moved for the small scale to the large scale. The volume of the water droplet is 20μL, and the frequency and the amplitude of the oscillator are 80 Hz and 1.5 mm, respectively.